\documentclass[aps,prl,twocolumn,numerical,superscriptaddress,nofootinbib,showpacs,showkeys,longbibliography]{revtex4-1}

\usepackage{bm}
\usepackage{graphicx,color}
\usepackage{amsmath,amssymb,amsfonts}
\usepackage[colorlinks=true,linkcolor=blue,plainpages=false]{hyperref}

\newcommand{\be}{\begin{eqnarray}}
\newcommand{\ee}{\end{eqnarray}}

\def\v2{\mbox{$v_2$}}

\def\eq{{\,=\,}}

\usepackage{ulem}
\usepackage{color}

\begin{document}


\title{Acoustic scaling of linear and mode-coupled anisotropic flow; 
implications for precision extraction of the specific shear viscosity}

\author{Peifeng Liu} 
\affiliation{Department of Chemistry, 
Stony Brook University, \\
Stony Brook, NY, 11794-3400, USA}
\affiliation{Department of Physics and Astronomy, Stony Brook University, \\
Stony Brook, NY, 11794-3800}
\author{ Roy~A.~Lacey}
\email[E-mail: ]{Roy.Lacey@Stonybrook.edu}
\affiliation{Department of Chemistry, 
Stony Brook University, \\
Stony Brook, NY, 11794-3400, USA}
\affiliation{Department of Physics and Astronomy, Stony Brook University, \\
Stony Brook, NY, 11794-3800}
%
%
%
%



\date{\today}

\begin{abstract}
The  $\mathrm{n^{th}}$-order linear flow coefficients $\mathrm{v^L_n \, (n=2,3,4,5)}$,
and the corresponding nonlinear mode-coupled ($\mathrm{mc}$) coefficients $\mathrm{v^{mc}_{4,(2,2)}}$, 
$\mathrm{v^{mc}_{5,(2,3)}}$, $\mathrm{v^{mc}_{6,(3,3)}}$ and $\mathrm{v^{mc}_{6,(2,2,2)}}$, 
are studied for Pb+Pb collisions at $\sqrt{s_{_{\rm NN}}} = 2.76$ TeV. 
Both sets of coefficients indicate a common acoustic scaling pattern of 
exponential viscous modulation, with a rate proportional to the square of the 
harmonic numbers and the mean transverse momenta (respectively), and inversely 
proportional to the cube root of the charge particle multiplicity ($\mathrm{(N_{ch})^{1/3}}$), 
that characterizes the dimensionless size of the systems produced in the collisions. 
These patterns and their associated scaling parameters, provide new stringent constraints for 
eccentricity independent estimates of the specific shear viscosity ($\eta/s$) and the   
viscous correction to the thermal distribution function for the matter produced in the collisions.
They also give crucial constraints for extraction of the initial-state eccentricity spectrum. 
\end{abstract}

\pacs{ }

\date{\today}

\maketitle


Anisotropic flow measurements play a crucial role in ongoing studies of the properties of the high 
energy-density quark-gluon plasma (QGP) created in relativistic heavy-ion 
collisions \cite{Teaney:2003kp,Lacey:2006pn,Romatschke:2007mq,Luzum:2008cw,Song:2010mg,Qian:2016fpi,Schenke:2011tv,Gardim:2012yp}. 
In particular, they provide an important avenue for the extraction of the specific shear viscosity 
(i.e., the ratio of shear viscosity to entropy density $\eta/s$) of the QGP, since they encode the 
viscous hydrodynamic response to the anisotropic transverse energy density profile 
produced in the early stages of the collision
\cite{Romatschke:2007mq,Luzum:2011mm,Song:2010mg,Qian:2016fpi,Schenke:2011tv,Teaney:2012ke,Gardim:2012yp}.

In experiments, this flow manifests as an azimuthal asymmetry of the measured single-particle distribution
and is routinely quantified by the complex flow vectors \cite{Ollitrault:1992bk,Luzum:2011mm,Teaney:2012ke}:
\begin{equation}
 \mathrm{V_n  \equiv v_ne^{in\Psi_n} \equiv \{e^{in\phi}\}}, \ \ \mathrm{v_n = {\left< \left| V_n \right|^2 \right>}^{1/2}},
\label{Vndef}
\end{equation}
where $\phi$ denotes the azimuthal angle around the beam direction, of a particle emitted 
in the collision, $\{\dots\}$ denotes the average over all particles emitted in the event, 
and $\mathrm{v_n}$ and $\mathrm{\Psi_n}$ denote 
the magnitude and azimuthal direction of the $\mathrm{n^{th}}$-order harmonic flow vector 
which fluctuates from event to event. The coefficients $\mathrm{v_2}$ and $\mathrm{v_3}$
are commonly termed elliptic- and triangular flow respectively.

The initial anisotropic density profile $\rho_e(r,\varphi)$ (in the transverse plane) which 
drives anisotropic flow, can be similarly characterized by complex eccentricity 
coefficients \cite{Alver:2010dn,Petersen:2010cw,Lacey:2010hw,Teaney:2010vd,Qiu:2011iv}:
\begin{eqnarray}
\mathrm{\mathcal{E}_n  \equiv \varepsilon_n e^{in\Phi_n} \equiv 
  - \frac{\int d^2r_\perp\, r^m\,e^{in\varphi}\, \rho_e(r,\varphi)}
           {\int d^2r_\perp\, r^m\,\rho_e(r,\varphi)}},                                                       
\label{epsdef1}
\end{eqnarray}
where $\mathrm{\varepsilon_n = {\left< \left| \mathcal{E}_n \right|^2 \right>}^{1/2}}$ and $\mathrm{\Phi_n}$ denote 
the magnitude and azimuthal direction of the $\mathrm{n^{th}}$-order eccentricity vector 
which also fluctuates from event to event;
$\mathrm{m\eq{n}}$ for $\mathrm{n{\geq\,}2}$ and $\mathrm{m\eq3}$ for 
$\mathrm{n\eq1}$ \cite{Teaney:2010vd,Bhalerao:2014xra,Yan:2015jma}. 

Theoretical investigations show that $\mathrm{v_n \propto \varepsilon_n}$ for  elliptic- and triangular 
flow ($\mathrm{n=2 \: and \: 3}$) \cite{Qiu:2011iv,Fu:2015wba,Niemi:2015qia,Noronha-Hostler:2015dbi},
albeit with a small anti-correlation between $\mathrm{v_2}$ and $\mathrm{v_3}$ \cite{Aad:2015lwa,ALICE:2016kpq},
which derives from an anti-correlation between $\varepsilon_2$ and $\varepsilon_3$  \cite{Lacey:2013eia};
the latter is more important for peripheral collisions.
Because the specific shear viscosity $\eta/s$, reduces the values of $\mathrm{v_n}$ and hence, 
the ratio $\mathrm{v_n/\varepsilon_n}$, viscous hydrodynamical model comparisons to this 
ratio (implicit and explicit) have been employed to estimate $\eta/s$  
\cite{Hirano:2005xf,Romatschke:2007mq,Song:2010mg,Schenke:2010rr,Bozek:2010wt,
Qiu:2011iv,Schenke:2011tv,Niemi:2012ry,Gardim:2012yp}.
Such estimates have indicated a small value (i.e.  1-3 times the lower conjectured 
bound of ${1}/{4\pi}$ \cite{Kovtun:2004de}), with substantial uncertainties of ${\cal O}(100\%)$, 
primarily due to the lack of constraints for $\mathrm{\varepsilon_n}$ and its 
fluctuations. Thus, there is a pressing need to develop new experimental constraints that can 
reduce this critical bottleneck for precision extraction of $\eta/s$. 

   The higher order flow coefficients for $\mathrm{n> 3}$, reflect a linear response 
related to $\mathrm{\varepsilon_{n}}$, as well as nonlinear mode-couplings  
derived from lower-order harmonics driven by eccentricities of the same 
harmonic order \cite{Teaney:2012ke,Bhalerao:2014xra,Yan:2015jma}:
\begin{eqnarray}
\mathrm{V_{4}} &=&   \mathrm{V_{4}^{L} + \chi^{mc}_{4,(2,2)} (V_{2})^2}, \label{eq:V4}\\
\mathrm{V_{5}} &=&   \mathrm{V_{5}^{L} + \chi^{mc}_{5,(2,3)} V_{2} \, V_{3}}, \label{eq:V5}\\
\mathrm{V_{6}} &=&   \mathrm{V_{6}^{L} + \chi^{mc}_{6,(2,2,2)} (V_{2})^3 + \chi^{mc}_{6,(3,3)}(V_{3})^2}, \label{eq:V6}\\
\mathrm{V_{7}} &=&   \mathrm{V_{7}^{L} + \chi^{mc}_{7,(2,2,3)} (V_{2})^2 V_{3} } \label{eq:V7},
\end{eqnarray}
%
%
where $\mathrm{\chi^{mc}_{n,(i,j)}}$ and $\mathrm{\chi^{mc}_{n,(i,i,j)}}$ ($\mathrm{i=2,\, j=2,3}$) are 
$\mathrm{n^{th}}$-order nonlinear mode-coupling coefficients. In Eqs.~\ref{eq:V6} and \ref{eq:V7} 
the nonlinear contributions are restricted to the two largest flow 
coefficients, $\mathrm{V_2}$ and $\mathrm{V_3}$~\cite{Teaney:2012ke,Yan:2015jma}.

If the linear and non-linear terms in Eqs.~\ref{eq:V4} - \ref{eq:V7} are uncorrelated,  
the mode-coupling coefficients can be expressed as~\cite{Teaney:2012ke,Yan:2015jma}:
\begin{eqnarray}
\mathrm{\chi^{mc}_{4,(2,2)}} &=& \frac{\mathrm{Re} \langle V_4 (V_2^*)^2 \rangle}{\langle v_2^4 \rangle},\quad
\mathrm{\chi^{mc}_{5,(2,3)}} = \frac{\mathrm{Re} \langle V_5 V_2^* V_3^* \rangle}{\langle v_2^2 v_3^2 \rangle},
\nonumber\\
\mathrm{\chi^{mc}_{6,(3,3)}} &=& \frac{\mathrm{Re} \langle V_6 (V_3^*)^2 \rangle}{\langle v_3^4 \rangle},\quad
\mathrm{\chi^{mc}_{6,(2,2,2)}} = \frac{\mathrm{Re} \langle V_6 (V_2^*)^3 \rangle}{\langle v_2^6 \rangle},
\nonumber\\
\mathrm{\chi^{mc}_{7,(2,2,3)}} &=& \frac{\mathrm{Re} \langle V_7 (V_2^*)^2 V_3^* \rangle}{\langle v_2^4 v_3^2 \rangle}.
\label{chi_mc}
\end{eqnarray}
For a given $\mathrm{p_T}$ and centrality selection, the magnitudes of the mode-coupled flow vectors 
can also be expressed in terms of the correlations of $\mathrm{V_{n}}$ with $\Psi_{2}$ 
and $\Psi_3$ to give \cite{Bhalerao:2013ina,Yan:2015jma}:
\begin{eqnarray}
\mathrm{v^{mc}_{4,(2,2)}} &=& \mathrm{\frac{\langle v_{4} v_{2}^{2} \cos (4\Psi_{4} - 4\Psi_{2}) \rangle} 
{\sqrt{\langle  v_{2}^{4} \rangle }}}  
                          \approx \mathrm{\langle v_{4} \cos (4\Psi_{4} - 4\Psi_{2}) \rangle }, \nonumber \\ 
\mathrm{v^{mc}_{5,(3,2)}} &=& \mathrm{\frac{\langle v_{5} v_{3} v_{2} \cos (5\Psi_{5} - 3\Psi_{3} - 2\Psi_{2}) \rangle} 
{\sqrt{\langle  v_{3}^{2} \, v_{2}^{2}  \rangle }}} \nonumber \\
             &\approx& \mathrm{\langle v_{5} \, \cos (5\Psi_{5} - 3\Psi_{3} - 2\Psi_{2})  \rangle}, \nonumber \\
\mathrm{v^{mc}_{6,(2,2,2)}} &=& \mathrm{\frac{\langle v_{6} \, v_{2}^{3} \, \cos (6\Psi_{6} - 6\Psi_{2}) \rangle} 
{\sqrt{\langle  v_{2}^{6} \rangle }}}
                          \approx  \mathrm{\langle v_{6} \cos (6\Psi_{6} - 6\Psi_{2}) \rangle}, \nonumber \\ 
\mathrm{v^{mc}_{6,(3,3)}} &=& \frac{\langle v_{6} v_{3}^{2} \cos (6\Psi_{6} - 6\Psi_{3}) \rangle}{\sqrt{\langle v_{3}^{4} \rangle }}
                          \approx \mathrm{\langle v_{6} \cos (6\Psi_{6} - 6\Psi_{3}) \rangle}, \nonumber \label{eq:V2nApsin2}
\end{eqnarray}
where the average in the numerator is an average over particles for a given $\mathrm{p_T}$ selection, for all the events 
in the chosen centrality range, and the average in the denominator is an average over events for the centrality 
selection. These expressions point to the important role of event-plane correlations for mode-coupling.
It is also straight forward to use Eqs.~\ref{eq:V4} - \ref{chi_mc} to evaluate the magnitude 
of the higher-order linear harmonic response:
\begin{eqnarray}
\mathrm{v_{4}^{L} = \sqrt{v_{4}^{\,2} - v_{4,(2,2)}^{\,2}} }, \quad 
\mathrm{v_{5}^{L} = \sqrt{v_{5}^{\,2} - v_{5,(3,2)}^{\,2}}}.
\label{eq:vnPsim}
\end{eqnarray}

Analogous to anisotropic flow, the complex eccentricity coefficients defined in Eq.~\ref{epsdef1}, can be used 
to determine the higher-order mixed-mode eccentricities:
\begin{eqnarray}
\mathrm{\varepsilon_n}                &=& \sqrt{\left< \left| \mathcal{E}_n \right|^2 \right>}, \quad
\mathrm{{\varepsilon}^{mc}_{4,(2,2)}}  = \sqrt{\langle \epsilon_2^4 \rangle}, \nonumber\\
\mathrm{{\varepsilon}^{mc}_{5,(2,3)}} &=& \sqrt{\langle \epsilon_2^2 \epsilon_3^2 \rangle}, \quad 
\mathrm{{\varepsilon}^{mc}_{6,(3,3)}}  = \sqrt{\langle \epsilon_3^4 \rangle}, \nonumber\\
\mathrm{{\varepsilon}^{mc}_{6,(2,2,2)}} &=& \sqrt{\langle\epsilon_2^6\rangle}, \quad 
\mathrm{{\varepsilon}^{mc}_{7,(2,2,3)}}  =  \sqrt{\langle\epsilon_2^4\epsilon_3^2 \rangle}. \label{epsdef2}
\end{eqnarray}
Recently, it has been argued that the linear response contribution to higher-order flow, 
should be linearly proportional to the cumulant-defined eccentricities $\mathrm{\mathcal{E}'_n}$ 
instead of $\mathrm{\mathcal{E}_n}$ \cite{Teaney:2012ke}: 
\begin{equation}
\begin{split}
 \mathcal{E}_2' &\equiv \epsilon_2 e^{i2\Phi_2} = \mathcal{E}_2, \quad \quad \quad 
  \mathcal{E}_3' \equiv \epsilon_3 e^{i3\Phi_3} = \mathcal{E}_3, \\
 \mathcal{E}_4' &\equiv \epsilon_4' e^{i4\Phi_4'} \equiv - \frac{\langle z^4 \rangle - 3 \langle z^2 \rangle ^2}{\langle r^4 \rangle}  = \mathcal{E}_4 + \frac{3\langle r^2 \rangle ^2}{\langle r^4 \rangle} \mathcal{E}_2^2, \\
 \mathcal{E}_5' &\equiv \epsilon_5' e^{i5\Phi_5'} \equiv - \frac{\langle z^5 \rangle - 10 \langle z^2 \rangle \langle z^3 \rangle}{\langle r^5 \rangle}  = \mathcal{E}_5 + \frac{10\langle r^2 \rangle \langle r^3 \rangle}{\langle r^5 \rangle} \mathcal{E}_2 \mathcal{E}_3,
\end{split}
\label{EcnNew}
\end{equation}
where $z \equiv x+iy = re^{i\phi}$.
An important advantage of this definition, is that it allows the subtraction of contributions from 
lower order $z$ correlations. 

In analogy to elliptic and triangular flow, $\mathrm{{v^{L}_n} \propto {\varepsilon^{'}_n}}$, 
$\mathrm{{v^{mc}_{n,(i,j)}} \propto {\varepsilon^{mc}_{n,(i,j)}}}$ and 
$\mathrm{{v^{mc}_{n,(i,i,j)}}\propto {\varepsilon^{mc}_{n,(i,i,j)}}}$. 
The specific shear viscosity also attenuates $\mathrm{{v^{L}_n}/{\varepsilon^{'}_n}}$, 
$\mathrm{{v^{mc}_{n,(i,j)}}/{\varepsilon^{mc}_{n,(i,j)}}}$ and 
$\mathrm{{v^{mc}_{n,(i,i,j)}}/{\varepsilon^{mc}_{n,(i,i,j)}}}$.
For measurements at a given mean transverse momentum $\mathrm{\left< p_T \right>}$, and centrality $\mathrm{cent}$, 
this viscous damping can be expressed via an 
acoustic ansatz \cite{Lacey:2011ug,Lacey:2013is,Shuryak:2013ke,Lacey:2013eia} as:
\begin{eqnarray}
\mathrm{\frac{v^{L}_n}{\varepsilon^{'}_n}} &\propto& \mathrm{\exp{\left(-n^2 \beta \frac{1}{{RT}} \right)}},
\label{a_damp-L} \\
\mathrm{\frac{v^{mc}_{n,(i,j)}}{\varepsilon^{mc}_{n,(i,j)}}} 
                     &\propto& \mathrm{\exp{\left(-(i^2 + j^2)\beta \frac{1}{{RT}} \right)} }, \nonumber \\
\mathrm{\frac{v^{mc}_{n,(i,i,j)}}{\varepsilon^{mc}_{n,(i,i,j)}}} 
                     &\propto& \mathrm{\exp{\left(-(2i^2 + j^2)\beta \frac{1}{{RT}} \right)} },										
\label{a_damp-mc}
\end{eqnarray}
where $\beta \propto {\eta}/{s}$, $\mathrm{T}$ is the temperature and $\mathrm{R}$ characterizes the  
geometric size of the collision zone. For a given centrality selection, the dimensionless 
size $\mathrm{RT \propto N_{ch}^{1/3}}$, where $\mathrm{N_{ch}}$ is the charged 
particle multiplicity density in one unit of pseudorapidity \cite{Lacey:2016hqy}.

Equations \ref{a_damp-L} and \ref{a_damp-mc} suggest characteristic linear 
dependencies for $\mathrm{\ln(v^{L}_n/\varepsilon^{'}_n)}$, $\mathrm{\ln(v^{mc}_{n,(i,j)}/\varepsilon^{mc}_{n,(i,j)})}$ 
and $\mathrm{\ln(v^{mc}_{n,(i,i,j)}/\varepsilon^{mc}_{n,(i,i,j)})}$ 
on $\mathrm{\left< N_{ch} \right>^{-1/3}}$ (respectively), 
with slopes that reflect specific quadratic viscous attenuation prefactors for $\beta$; these combined features are 
termed acoustic scaling. 
The prefactors, reflected in the slopes of  
$\mathrm{\ln(v^{L}_n/\varepsilon^{'}_n)}$ vs. $\mathrm{(N_{ch})^{-1/3}}$, 
are not only expected to increase as $\mathrm{n^2}$, but should be approximately 2-3 times larger 
than those for $\mathrm{\ln(v^{mc}_{n,(i,j)}/\varepsilon^{mc}_{n,(i,j)})}$ and 
$\mathrm{\ln(v^{mc}_{n,(i,i,j)}/\varepsilon^{mc}_{n,(i,i,j)})}$ vs. $\mathrm{(N_{ch})^{-1/3}}$ (respectively)
since $\mathrm{(i^2 + j^2) < n^2}$.

Independent estimates of $\beta$, involving very different eccentricities, can also be obtained 
from the linear and mode-coupled harmonics. For example, the slope of the  
double ratio $\ln[(\mathrm{v^{mc}_{5,(2,3)}/\varepsilon^{mc}_{5,(2,3)}})/(\mathrm{v_2/\varepsilon_2})]$
vs. $\mathrm{(N_{ch})^{-1/3}}$, is expected to be similar to that for   
$\mathrm{\ln(v_3/\varepsilon_3)}$ vs. $\mathrm{(N_{ch})^{-1/3}}$ for a given $\mathrm{\left< p_T \right>}$. 
Thus, the validation of simultaneous acoustic scaling of the linear and mode-coupled harmonics to give 
a single estimate of $\beta \propto \eta/s$, could provide a powerful constraint for 
initial-state eccentricity models and precision extraction of $\eta/s$. 

In this letter, we use recent measurements of the linear and mode-coupled harmonics in Pb+Pb collisions at 
$\sqrt{s_{NN}}$ = 2.76 TeV, to explore validation tests for simultaneous 
acoustic scaling of $\mathrm{v^{L}_n/\varepsilon^{'}_n}$, $\mathrm{v^{mc}_{n,(i,j)}/\varepsilon^{mc}_{n,(i,j)}}$ 
and $\mathrm{v^{mc}_{n,(i,i,j)}/\varepsilon^{mc}_{n,(i,i,j)}}$, with an eye towards the 
development of new experimental constraints which could significantly reduce the large 
eccentricity-driven uncertainties associated with current extractions of $\eta/s$.

The data employed in this work are taken from the published flow measurements 
for Pb+Pb collisions at $\sqrt{s_{NN}}$ = 2.76 TeV by the ALICE \cite{Acharya:2017zfg,ALICE:2011ab} 
and ATLAS \cite{Aad:2015lwa} collaborations.
The ALICE centrality dependent $\mathrm{p_T}$-integrated measurements were performed for the harmonics $\mathrm{n = 2,3,4,5,6}$, for
charged particles with pseudorapidity difference $\mathrm{|\Delta\eta| < 0.8}$ and $\mathrm{0.2 < p_T < 5.0}$ GeV/c. 
Both the linear and mode-coupled flow coefficients were obtained directly via a two sub-events multiparticle 
correlation method.
The corresponding ATLAS measurements were performed for $\mathrm{n = 2,3,4,5}$ for
particles with $2<|\Delta\eta| <5$ and for several $\mathrm{p_T}$ selections spanning the range $\mathrm{0.5 < p_T < 4.0}$ GeV/c,
with the two-particle correlation method supplemented with event-shape selection \cite{Aad:2015lwa}.
The systematic uncertainties, which are included in our scaling analyses, 
are reported in Refs.~\cite{Acharya:2017zfg,ALICE:2011ab,Aad:2015lwa} for both sets of measurements.
%
\begin{figure}
\includegraphics[width=1.05\linewidth]{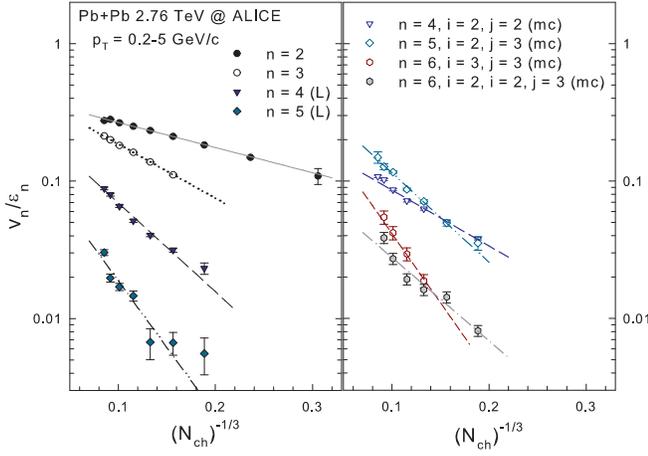}
\caption{Comparison of $\mathrm{(v^{L}_n/\varepsilon^{'}_n)}$ vs. $\mathrm{(N_{ch})^{-1/3}}$ 
for the linear harmonics (left panel), and $\mathrm{v^{mc}_{n,(i,j)}/\varepsilon^{mc}_{n,(i,j)}}$ and 
$\mathrm{v^{mc}_{n,(i,i,j)}/\varepsilon^{mc}_{n,(i,i,j)}}$ vs. $\mathrm{(N_{ch})^{-1/3}}$ (respectively)
for the nonlinear mode-coupled harmonics, for Pb+Pb collisions at at $\sqrt{s_{NN}}$ = 2.76 TeV. 
The lines represent a simultaneous exponential fit to the data, following Eqs.~\ref{a_damp-L} and \ref{a_damp-mc}.
The ALICE data are taken from Refs.~\cite{Acharya:2017zfg,ALICE:2011ab}.
}
\label{fig:1}
\end{figure} 
%

The requisite cumulant-defined eccentricities were calculated following the procedure 
outlined in Eqs.~\ref{epsdef1}, \ref{epsdef2} and \ref{EcnNew}
with the aid of a Monte Carlo quark-Glauber model (MC-qGlauber) with fluctuating initial conditions \cite{Pifeng-Liu}.
The model, which is based on the commonly used MC-Glauber model \cite{Miller:2007ri,*Alver:2006wh}, 
was used to compute the number of quark participants $\mathrm{Nq_{part}(cent)}$, and 
$\mathrm{\varepsilon^{'}_n(cent)}$ and $\mathrm{\varepsilon^{mc}_n(cent)}$ from the two-dimensional profile of the 
density of sources in the transverse  plane $\rho_s(\mathbf{r_{\perp}})$ \cite{Pifeng-Liu,Lacey:2010hw,Teaney:2012ke}. 
The model takes account of the finite size of the nucleon, the wounding profile of the nucleon,
the distribution of quarks inside the nucleon and quark cross sections which reproduce 
the NN inelastic cross section at $\sqrt{s_{NN}}$ = 2.76 TeV. A systematic uncertainty 
of 2-5\% was estimated for the eccentricities from variations of the model parameters.

The centrality dependent multiplicity densities used to evaluate the dimensionless 
size $\mathrm{RT \propto N_{ch}^{1/3}}$, are obtained 
from ALICE \cite{Aamodt:2010cz} and ATLAS \cite{ATLAS:2011ag} 
multiplicity density measurements. Validation tests for acoustic scaling were performed 
by plotting $\mathrm{v^{L}_n/\varepsilon^{'}_n}$, $\mathrm{v^{mc}_{n,(i,j)}/\varepsilon^{mc}_{n,(i,j)}}$ 
and $\mathrm{v^{mc}_{n,(i,i,j)}/\varepsilon^{mc}_{n,(i,i,j)}}$ vs. $\mathrm{\left< N_{ch} \right>^{-1/3}}$
respectively, to test for the expected patterns of exponential viscous attenuation, and the relative 
viscous attenuation $\beta$-prefactors indicated in Eqs.~\ref{a_damp-L} and \ref{a_damp-mc}.

\begin{figure}
\includegraphics[width=1.05\linewidth]{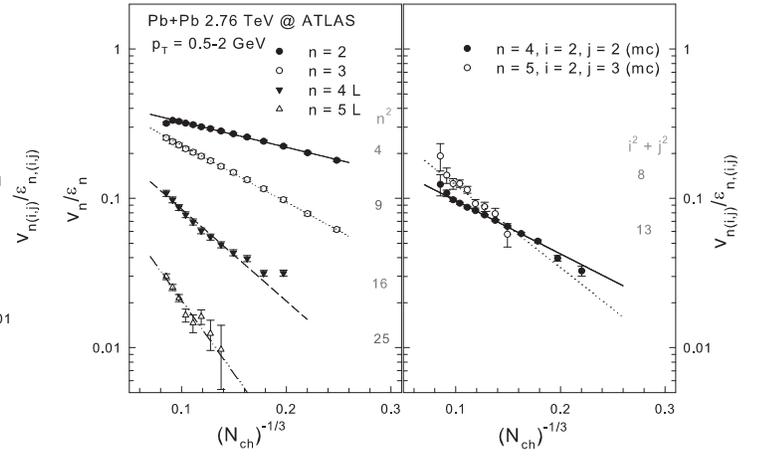}
\caption{Same as Fig.~\ref{fig:1} but for ATLAS data \cite{Aad:2015lwa};
the $\beta$-prefactors $\mathrm{n^2}$ and $\mathrm{(i^2 + j^2)}$ are indicated 
in the figure.
}
\label{fig:2}
\end{figure}

Figures \ref{fig:1} and \ref{fig:2} show the plots for $\mathrm{v^{L}_n/\varepsilon^{'}_n}$, 
$\mathrm{v^{mc}_{n,(i,j)}/\varepsilon^{mc}_{n,(i,j)}}$ 
and $\mathrm{v^{mc}_{n,(i,i,j)}/\varepsilon^{mc}_{n,(i,i,j)}}$ vs. $\mathrm{(N_{ch})^{-1/3}}$
(respectively), for the ALICE (Fig.~\ref{fig:1}) and ATLAS (Fig.~\ref{fig:2}) data sets. 
They indicate the telltale acoustic scaling patterns of a characteristic linear dependence 
of $\mathrm{\ln(v^{L}_n/\varepsilon^{'}_n)}$, 
$\mathrm{\ln(v^{mc}_{n,(i,j)}/\varepsilon^{mc}_{n,(i,j)})}$ 
and $\mathrm{\ln(v^{mc}_{n,(i,i,j)}/\varepsilon^{mc}_{n,(i,i,j)})}$
on $\mathrm{(N_{ch})^{-1/3}}$ (respectively), with slope factors which strongly depend on the harmonic
number $\mathrm{n}$ and the values of the mode-coupled harmonics $\mathrm{i,j}$ and $\mathrm{i,i,j}$.
Note that the slopes for the linear harmonics (left panel in each figure) show a much steeper dependence  
on $\mathrm{(N_{ch})^{-1/3}}$ than those for the mode-coupled harmonics (right panel in each figure), as 
expected from Eqs.~\ref{a_damp-L} and \ref{a_damp-mc}. The expected slope hierarchy 
for both the linear and mode-coupled results are also apparent 
in both figures. The qualitative similarities between the results shown in Figs. \ref{fig:1} and \ref{fig:2}
suggest that the respective methods employed by ATLAS and ALICE for extraction of the 
flow coefficients, are complementary.
%
%

%
\begin{figure*}
\includegraphics[width=0.75\linewidth]{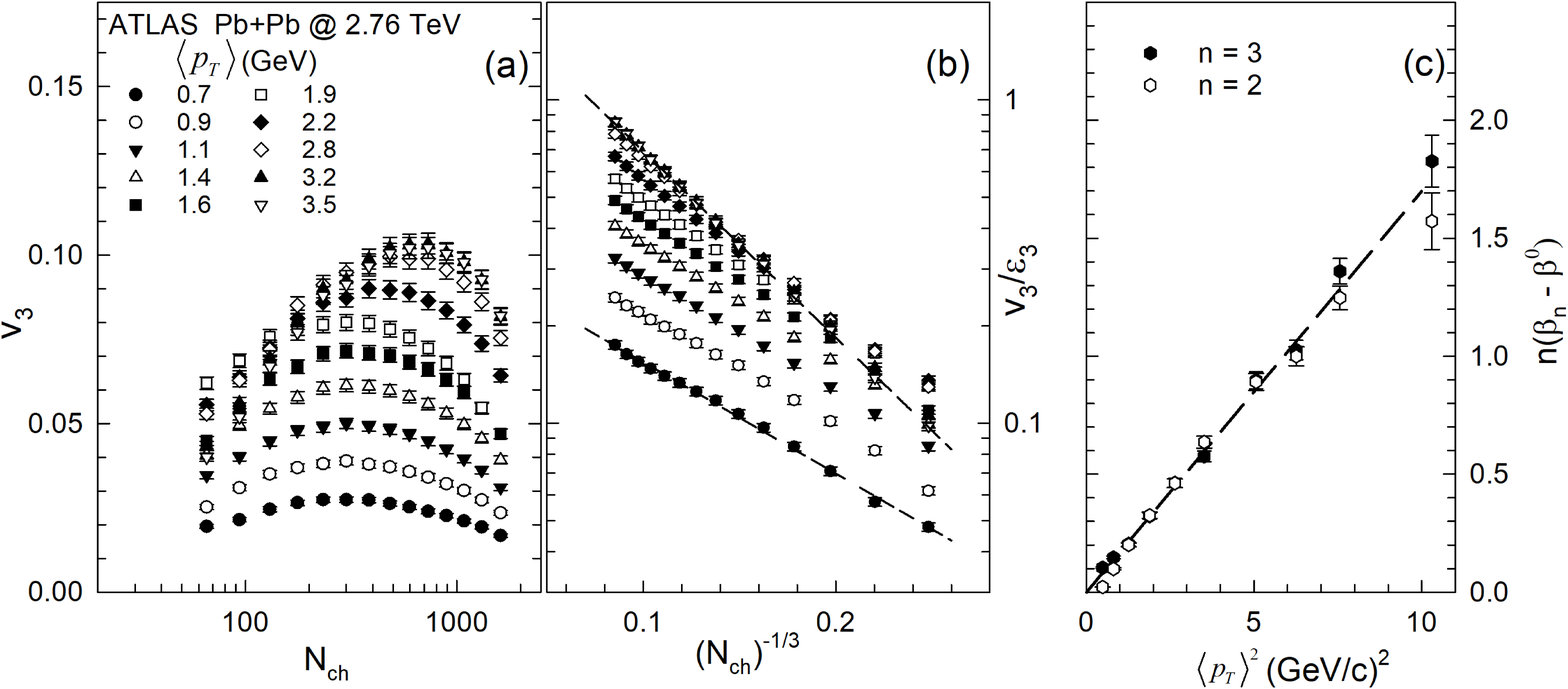}
\caption{(a) $\mathrm{v_3}$ vs. $\mathrm{N_{ch}}$ for several $\mathrm{\left< p_T \right>}$ selections as indicated; 
(b) $\mathrm{v_3/\varepsilon_3}$ vs. $\mathrm{(N_{ch})^{-1/3}}$ for the data 
shown in (a). The dashed lines represent an exponential fit to the data for the selections
$\mathrm{\left< p_T \right> =}$~0.7 and 3.5 GeV/c respectively. 
(c) $\mathrm{n(\beta_n -\beta^0)}$ vs. $\mathrm{\left< p_T \right>^2}$ (see text); 
the slopes $\mathrm{n\beta_n}$, are obtained from fits to the eccentricity-scaled data, similar to that 
shown in (b). The ATLAS data used in the plots are taken from Ref.~\cite{Aad:2015lwa}
}
\label{fig:3}
\end{figure*} 
%

The lines shown in Figs. \ref{fig:1} and \ref{fig:2} represent the results from fits to the 
data following Eqs.~\ref{a_damp-L} and \ref{a_damp-mc}. They indicate that, within an uncertainty of 
$\sim 2-12$\%, a single slope value $\beta$, can account for the wealth of the linear and 
mode-coupled measurements in each data set. 
That is, they confirm the quadratic $\beta$ prefactors of 4, 9, 16 and 25 for 
$\mathrm{v^{L}_n}$ (n=2,3,4 and 5) and 8, 13, 18 and 12 for $\mathrm{v^{mc}_{4,(2,2)}}$, 
$\mathrm{v^{mc}_{5,(2,3)}}$, $\mathrm{v^{mc}_{6,(3,3)}}$ and $\mathrm{v^{mc}_{6,(2,2,2)}}$ 
respectively. To estimate the fit uncertainty for each data set, the slope for the fit 
to $\mathrm{v_2/\varepsilon_2}$ was first obtained, and then used in conjunction with 
the quadratic prefactors to quantify slope deviations from one. 

The value of $\beta$ also depend on $\mathrm{p_T}$, even though this is not explicitly indicated 
in Eqs.~\ref{a_damp-L} and \ref{a_damp-mc}. In hydrodynamical models, this $\mathrm{p_T}$ dependence can 
be understood in terms of the first viscous correction $\delta f$, to the thermal distribution 
function \cite{Dusling:2009df,Teaney:2009qa}. It leads to an additional viscous attenuation factor
$\mathrm{\propto p_T^{\alpha}}$, where current theoretical estimates indicate the range 1-2 
for ${\alpha}$ \cite{Dusling:2009df,Teaney:2009qa}.
That is, $\beta$ is expected to increase as $\mathrm{p_T^{\alpha}}$, where the value of ${\alpha}$ is currently 
not fully constrained. 

An experimental constraint for $\alpha$ can be obtained via acoustic scaling of the differential 
measurements $\mathrm{v_n(N_{ch})}$, for different $\mathrm{\left< p_T \right>}$ selections as illustrated 
in Fig. \ref{fig:3}. Panel (a) shows a steepened decrease of $\mathrm{v_3}$ with $\mathrm{\left< p_T \right>}$, 
for $\mathrm{N_{ch} \alt 400}$. This pattern results from an increase in the viscous attenuation with 
$\mathrm{\left< p_T \right>}$. This attenuation is made more transparent in Fig. \ref{fig:3}(b), 
where $\mathrm{(v_3/\varepsilon_3)}$ vs. $\mathrm{(N_{ch})^{-1/3}}$
is plotted for several $\mathrm{\left< p_T \right>}$ selections as indicated. The characteristic linear dependence 
of $\mathrm{\ln(v_3/\varepsilon_3)}$ on $\mathrm{(N_{ch})^{-1/3}}$ (i.e., exponential viscous 
attenuation), is clearly visible for each $\mathrm{\left< p_T \right>}$ selection. It is also apparent 
that the slopes $\beta$, for $\mathrm{\ln(v_3/\varepsilon_3)}$ vs. $\mathrm{(N_{ch})^{-1/3}}$ increases with  
$\mathrm{\left< p_T \right>}$ over the range indicated. This increase reflects the additional viscous 
attenuation factor due to $\delta f$.

The slopes, obtained from fits to $\mathrm{(v_3/\varepsilon_3)}$ vs. $\mathrm{(N_{ch})^{-1/3}}$ 
(c.f. panel (b)) and $\mathrm{(v_2/\varepsilon_2)}$ vs. $\mathrm{(N_{ch})^{-1/3}}$, for 
each $\mathrm{\left< p_T \right>}$ selection, are plotted vs. $\mathrm{\left< p_T \right>^2}$ in panel (c). 
Note that the plotted slopes are $\mathrm{\beta^{\delta f}_n \equiv n(\beta_n - \beta^0)}$, 
where $\mathrm{\beta^0 = 0.83 \pm 0.04}$,
is the value for $\mathrm{p_T = 0.0}$ GeV/c. The dashed line, which shows a linear fit to the data, 
indicates that $\mathrm{\beta^{\delta f}_n}$ increases as $\mathrm{\left< p_T \right>^2}$, 
i.e., $\mathrm{\beta^{\delta f}_n = nkp_T^2}$ where $\mathrm{k = 0.169 \pm 0.003 \, GeV^{-2}}$ for 
these data. These results provide a clear constraint for $\alpha$ and $\mathrm{\beta^{\delta f}_n}$, 
and consequently, the first viscous correction to the thermal distribution function in 
viscous hydrodynamical models.

\begin{figure}
\includegraphics[width=1.0\linewidth]{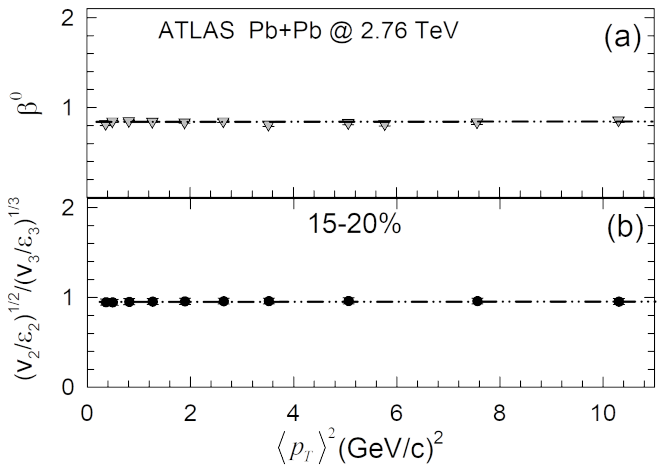}
\caption{ (a) $\beta^0$ vs. $\mathrm{\left< p_T \right>^2}$; the $\beta^0$ values are extracted from 
plots of $\mathrm{\ln[(v_3/\varepsilon_3)/(v_2/\varepsilon_2)]}$ vs. $\mathrm{(N_{ch})^{-1/3}}$,
for several $\mathrm{\left< p_T \right>}$ selections (see text).
(b) $(v_2/\varepsilon_2)^{1/2}/(v_3/\varepsilon_3)^{1/3}$ vs. $\mathrm{\left< p_T \right>^2}$ for 
15-20\% central Pb+Pb collisions. The dashed lines in both panels are drawn to guide the eye.
The ATLAS data used in the plots are taken from Ref.~\cite{Aad:2015lwa}
}
\label{fig:4}
\end{figure}

The scaling patterns shown in Fig.~\ref{fig:3}(c) indicate that the viscous coefficient in 
Eq.~\ref{a_damp-L} can be expressed as $\mathrm{n^2\beta = n(n\beta^0 + kp_T^2)}$ and used to extract 
$\beta^0$ from ratios of the eccentricity scaled harmonics. Fig. \ref{fig:4}(a) shows the $\beta^0$ 
values extracted from 
$\mathrm{\ln[(v_3/\varepsilon_3)/(v_2/\varepsilon_2)]}$ vs. $\mathrm{(N_{ch})^{-1/3}}$ for 
several values of $\mathrm{\left< p_T \right>}$; the prefactors are 5 ($\mathrm{n^2-m^2}$) and 1 ($\mathrm{n-m}$) for 
$\beta^0$  and $\mathrm{\beta^{\delta f}_{n-m}}$, respectively.
Fig. \ref{fig:4}(a) indicates that the extracted $\beta^0$ values are essentially $\mathrm{p_T}$-independent 
over the $\mathrm{\left< p_T \right>}$ range of interest. This $\mathrm{p_T}$-independence confirms 
that the pattern of viscous attenuation, due to $\delta f$,
is similar for $\mathrm{v_n}$ and $\mathrm{v_m}$ with magnitudes that differ by the value $\mathrm{(n-m)kp_T^2}$.  
Fig. \ref{fig:4}(b) shows that similar magnitudes and trends are obtained for the empirical ratio
$(v_2/\varepsilon_2)^{1/2}/(v_3/\varepsilon_3)^{1/3}$ vs. $\mathrm{\left< p_T \right>^2}$ \cite{Lacey:2011ug}, 
indicating that the $\delta f$-driven viscous attenuation factor $\mathrm{nkp_T^2}$, cancels for this ratio.
Thus, the ratio $(v_n/\varepsilon^{'}_n)^{1/n}/(v_2/\varepsilon_2)^{1/2}$ vs. $\mathrm{\left< p_T \right>^2}$
can be used to further constrain $\beta^0$ and the eccentricity spectrum.

The present analysis shows that an eccentricity- and $\mathrm{p_T}$-independent estimate of $\beta^{0} \propto \eta/s$ 
can be constrained by simultaneous acoustic scaling of both the linear and 
mode-coupled differential flow coefficients. However, a further calibration would be required to map 
$\beta^{0}$ on to the the actual value of $\eta/s$ for the QGP. An appropriately 
constrained set of viscous hydrodynamical calculations, tuned to reproduce the results shown in 
Figs. \ref{fig:1} - \ref{fig:4}, could provide such a calibration to give a relatively precise 
estimate of $\eta/s$, as well as simultaneous verification of the initial-state 
eccentricity spectrum.
%


In summary, we have presented a detailed phenomenological investigation for new constraints 
designed to facilitate precision extraction of $\eta/s$. We find that the 
linear flow coefficients $\mathrm{v^L_n \, (n=2,3,4,5)}$, and the nonlinear mode-coupled 
coefficients $\mathrm{v^{mc}_{4,(2,2)}}$, $\mathrm{v^{mc}_{5,(2,3)}}$, 
$\mathrm{v^{mc}_{6,(3,3)}}$ and $\mathrm{v^{mc}_{6,(2,2,2)}}$, 
follow a common acoustic scaling pattern of exponential viscous modulation 
in the created medium, at a rate proportional to the square of the 
harmonic numbers, and inversely proportional to the dimensionless size $\mathrm{RT \propto (N_{ch})^{1/3}}$. 
The scaling patterns of specific ratios of the eccentricity scaled harmonics, also indicate
a characteristic square dependence on particle transverse momenta. 
These patterns and their associated scaling parameters, could provide stringent new constraints 
for eccentricity independent 
estimates of $\eta/s$ and the first viscous correction to the thermal distribution function, 
as well as the initial-state eccentricity spectrum.

{\bf Acknowledgments}
This research is supported by the US DOE under contract DE-FG02-87ER40331.A008. 
 




%
\bibliography{mode_coupled_vn} 
\end{document}